\documentclass[preprint2]{aastex}

\usepackage{epsf}
\usepackage{graphics}

\newcommand{\kms}{km s$^{-1}$}
\newcommand{\cts}{cts s$^{-1}$}
\newcommand{\ax}{$\alpha_{\rm X}$}
\newcommand{\aox}{$\alpha_{\rm ox}$}
\newcommand{\cm}{cm$^{-2}$}
\newcommand{\rb}[1]{\raisebox{1.5ex}[-1.5ex]{#1}}

\newcommand{\pl}{$\pm$}
\newcommand{\nh}{$N_{\rm H}$}
\shorttitle{Mkn 1239}
\shortauthors{Grupe, Mathur & Komossa}

\begin{document}

\def\clipfig#1{\def\lbracket{[}\def\testit{#1}%
    \ifx\testit\lbracket\let\next=\optclipfig\else\let\next=\stdclipfig\fi%
    \next{#1}}
%
%                  -- for backwards compatibility --
\newcommand {\hclipfig} [7] {\clipfig[#7]{#1}{#2}{#3}{#4}{#5}{#6}}
%
%       -- the mode select macros (PSFIG- or EPSF-mode) --
\def\usemodepsfig {\global\def\cfmode{x}\typeout{*** set clipfig to PSFIG mode ***}}
\def\usemodeepsf  {\global\def\cfmode{}\typeout{*** set clipfig to EPSF mode ***}}
\def\useunitmm    {\global\def\cfunit{x}\typeout{*** set clipfig to use mm as unit ***}}
\def\useunitcm    {\global\def\cfunit{}\typeout{*** set clipfig to use cm as unit ***}}
\def\clipfigsettings {\ifx\cfmode\empty\def\ccfmode{EPSF }\else\def\ccfmode{PSFIG }\fi%
    \ifx\cfunit\empty\def\ccfunit{cm }\else\def\ccfunit{mm }\fi%
    \typeout{*** current clipfig settings: \ccfmode mode, using \ccfunit as unit ***}}
%
%
%============ end of public macros ========================================
%
%============ macros for internal use =====================================
%
\def\stdclipfig#1#2#3#4#5#6{\ifx\cfmode\empty%
    \let\next=\eclipfig\else\let\next=\pclipfig\fi%
    \next{#1}{#2}{#3}{#4}{#5}{#6}}
\def\optclipfig#1#2]#3#4#5#6#7#8{\ifx\cfmode\empty%
    \let\next=\ehclipfig\else\let\next=\phclipfig\fi%
    \next{#3}{#4}{#5}{#6}{#7}{#8}{#2}}
%
%==========================================================================
%
%  These two macros are the PSFIG implementation (which are quite simple
%     because PSFIG accepts - beside others - dimensions given in units of cm,
%     which is exactly what we want here).
%
\newcommand {\pclipfig}[6] {\ifx\cfunit\empty%
        \psfig{figure=#1.ps,width=#2cm,bbllx=#3cm,bblly=#4cm,bburx=#5cm,%
           bbury=#6cm,clip=}\else%
        \psfig{figure=#1.ps,width=#2mm,bbllx=#3mm,bblly=#4mm,bburx=#5mm,%
           bbury=#6mm,clip=}\fi}
\newcommand {\phclipfig}[7] {\ifx\cfunit\empty%
        \hspace{#7cm}\psfig{figure=#1.ps,width=#2cm,bbllx=#3cm,bblly=#4cm,%
           bburx=#5cm,bbury=#6cm,clip=}\else%
        \hspace{#7mm}\psfig{figure=#1.ps,width=#2mm,bbllx=#3mm,bblly=#4mm,%
           bburx=#5mm,bbury=#6mm,clip=}\fi}
%
%==========================================================================
%
%  These two macros are the EPSF implementation (which look rather complicated
%     because EPSF expects the bounding box to be given in Postscript points
%     rather than in cm).
%
\newcommand {\eclipfig}[6]{%
  \ifx\cfunit\empty\epsfxsize=#2cm\else\epsfxsize=#2mm\fi%
  \epsfclipon\epsfverbosetrue%
  \cfcmtopspts{#3}\cfllxi=\cftempi\cfllxf=\cftempf%
  \cfcmtopspts{#4}\cfllyi=\cftempi\cfllyf=\cftempf%
  \cfcmtopspts{#5}\cfurxi=\cftempi\cfurxf=\cftempf%
  \cfcmtopspts{#6}\cfuryi=\cftempi\cfuryf=\cftempf%
  \def\cfstra{\number\cfllxi.\number\cfllxf}%
  \def\cfstrb{\number\cfllyi.\number\cfllyf}%
  \def\cfstrc{\number\cfurxi.\number\cfurxf}%
  \def\cfstrd{\number\cfuryi.\number\cfuryf}%
  \hbox{\epsfbox[{\cfstra} {\cfstrb} {\cfstrc} {\cfstrd}]{#1.ps}}}
\newcommand {\ehclipfig}[7]{%
  \ifx\cfunit\empty\epsfxsize=#2cm\else\epsfxsize=#2mm\fi%
  \epsfclipon\epsfverbosetrue%
  \cfcmtopspts{#3}\cfllxi=\cftempi\cfllxf=\cftempf%
  \cfcmtopspts{#4}\cfllyi=\cftempi\cfllyf=\cftempf%
  \cfcmtopspts{#5}\cfurxi=\cftempi\cfurxf=\cftempf%
  \cfcmtopspts{#6}\cfuryi=\cftempi\cfuryf=\cftempf%
  \def\cfstra{\number\cfllxi.\number\cfllxf}%
  \def\cfstrb{\number\cfllyi.\number\cfllyf}%
  \def\cfstrc{\number\cfurxi.\number\cfurxf}%
  \def\cfstrd{\number\cfuryi.\number\cfuryf}%
  \ifx\cfunit\empty\hspace{#7cm}\else\hspace{#7mm}\fi%
  \hbox{\epsfbox[{\cfstra} {\cfstrb} {\cfstrc} {\cfstrd}]{#1.ps}}%
  \vspace{-1mm}}
%
%==========================================================================
%
%  The following are some new TeX dimension identifiers for the
%  conversion of natural units (cm) into Postscript points (bp) (see below).
%
\newdimen\cfllxi \newdimen\cfllyi  \newdimen\cfurxi  \newdimen\cfuryi
\newdimen\cfllxf \newdimen\cfllyf  \newdimen\cfurxf  \newdimen\cfuryf
\newdimen\cftemp \newdimen\cftempi \newdimen\cftempf
\newdimen\cfpspoint \cfpspoint=1bp
%
%==========================================================================
%
%  Though LaTeX normally is not able to do arithmetics, these nifty macros
%  *do* arithmetics, namely the cm/mm-to-bp conversion (weird!).
%
\newcommand{\cfcmtopspts}[1]{\ifx\cfunit\empty%
  \cftemp=#1cm\else\cftemp=#1mm\fi%
  \multiply\cftemp10\divide\cftemp\cfpspoint%
  \cftempf=\cftemp\divide\cftemp10\cftempi=\cftemp\multiply\cftemp10%
  \advance\cftempf-\cftemp}
%
%=========================================================================
%
%  -- set and display startup mode and unit: --
\def\cfmode{}\def\cfunit{}\clipfigsettings
%
%  -- Il fine --

\useunitmm

\def \charthoffset {\hspace{0.2cm}} \def \charthsep {\hspace{0.3cm}}
\def \chartvsepcap {\vspace{0.3cm}}
\def \chartvsep {\vspace{0.1cm}}
\newcommand{\putcharta}[1]{\clipfig{#1}{89}{7}{5}{275}{185}}
\newcommand{\putchartb}[1]{\clipfig{#1}{82}{5}{3}{275}{185}}
\newcommand{\chartlineb}[2]{\parbox[t]{18cm}{\noindent\charthoffset\putcharta{#1}\charthsep\putchartb{#2}\chartvsep}}


\title{Mkn 1239: A highly polarized NLS1 with a steep  X-ray spectrum and strong
NeIX emission
\thanks{Based on
observations obtained with XMM-Newton, an ESA science mission with instruments
and contribution directly funded by ESA member states and NASA}
}


\author{D. Grupe\thanks{Guest observer, McDonald Observatory,
University of Texas at Austin} 
 \,and S. Mathur}
\affil{Astronomy Department, Ohio State University,
    140 W. 18th Ave., Columbus, OH-43210, U.S.A.}

\email{dgrupe, smita @astronomy.ohio-state.edu}

\author{S. Komossa}
\affil{MPI f\"ur extraterrestrische Physik, Giessenbachstr. 1, D-85748 Garching,
Germany}

\email{skomossa@mpe.mpg.de}


%\altaffiltext{1}{Visiting Astronomer, Cerro Tololo Inter-American Observat}


\begin{abstract}
We report the results of an XMM-Newton observation of the Narrow-Line
Seyfert 1 galaxy Mkn 1239. This optically highly polarized AGN has one of the
steepest X-ray spectra found in AGN with \ax=+3.0 based on ROSAT PSPC data.
The XMM-Newton EPIC PN and MOS data confirm this steep X-ray spectrum. 
The PN data are best-fit by a powerlaw with a partial covering
absorption model 
suggesting two light paths between the continuum source and the observer, one
indirect scattered one which is less absorbed and a highly absorbed direct light
path. This result agrees with the wavelength dependent degree of polarization in
the optical/UV band. 
Residuals in the X-ray
spectra of all three XMM-Newton EPIC detectors around 0.9 keV suggest
the presence of an emission line feature, most likely the Ne IX triplet.
The detection of NeIX and the non-detection of OVII/OVIII suggest a super-solar
Ne/O ratio.
\end{abstract}


\section{Introduction}

With the launch of the X-ray satellite ROSAT \citep{tru83} the X-ray energy
range down to 0.1 keV became accessible for the first time. During the half year
ROSAT All-Sky Survey (RASS, \citet{vog99})
 a large number of sources with steep X-ray
spectra were detected (\citet{tho98, beu99, schwo00}). 
About one third to one half
of these sources are AGN. \citet{gru96} and \citet{gru98a, gru03a} 
found that about 50\% of
bright soft X-ray selected AGN are Narrow-Line Seyfert 1 galaxies (NLS1s, 
\citet{oster85, good89}). They turned out to be the class of AGN with the steepest X-ray
spectra (e.g. \citet{bol96, gru96, gru98a, gru01, vau01, gru03a, wil02}). 
NLS1s are AGN with extreme properties which seem to be linked to each other:
An
increase in their X-ray spectral index \ax~correlates with the strength of the
optical FeII emission and anti-correlates with the widths of the Broad Line
Region (BLR) Balmer lines and the strength of the Narrow-Line Region (NLR) 
forbidden lines 
(e.g. \citet{gru96, gru99, gru03b, laor94, laor97, sul00}). 
All these relationships are governed by one fundamental underlying parameter,
usually called the \citet{bor92} 'Eigenvector-1' relation in AGN. The most
accepted explanation for Eigenvector 1 is the Eddington ratio $L/L_{Edd}$
\citep{bor02, sul00, gru03b, yuan03} in which NLS1s are AGN with the highest
Eddington ratios.

In a spectropolarimetry study of 18 NLS1s \citet{good89} found three
sources to show significant polarization that allowed a detailed
spectropolarimetric analysis: Mkn 766, Mkn 1239, and IRAS 1509--211. All these
sources show an increase of the degree of polarization towards the blue. In the
Broad Line Region
(BLR) Balmer lines the degree of polarization is larger then in the
continuum, while is is less in the Narrow Line Region (NLR) 
forbidden lines, suggesting the
scattering medium is 
somewhere located between the BLR and NLR (e.g. \citet{wills92}).
All three sources were
observed with the Position Sensitive Proportional Counter (PSPC, \citet{pfe86})
on board ROSAT (\citet{rush96a, gru98b, pfe01}). 
The NLS1 Mkn 1239 ($\alpha_{2000}$=09h 52m 19.1s;
$\delta_{2000}=-01^\circ~36^{'}~43^{''}$; z=0.019) 
had an unusually steep  X-ray spectral index \ax=2.9 \citep{bol96} during 
the ROSAT pointed observation.
Here we present the results of a serendipitous observation of Mkn 1239 with
XMM-Newton \citep{jan01}.

The outline of this paper is as follows: in \S\,\ref{observe} we describe the
XMM and ROSAT observations and the data reduction, in \S\,\ref{results} we
present the results of the ROSAT and XMM data analysis,
and in \S\,\ref{discuss} we discuss the results. 
Throughout the paper spectral indexes are denoted as energy spectral indexes
with
$F_{\nu} \propto \nu^{-\alpha}$. Luminosities are calculated assuming a Hubble
constant of $H_0$ =75 \kms Mpc$^{-1}$ and a deceleration parameter of $q_0$ =
0.0.

\section{\label{observe} Observations}

\subsection{XMM-Newton Observation}

Mkn 1239 was observed  by XMM-Newton during orbit 353 on
2001-11-12 from UT 20:04 - 22:57 for 5 ks with the EPIC PN (\citet{str01})
and 9.4 ks with the EPIC MOS (\citet{tur01}) with the thin filters in Full Frame
Mode. The background during the observation was low so all of the
observing times were used.

Source photons were collected in a circle with a radius of 25$^{''}$. The
background photons of the PN observation were collected in a circular region
close by with a radius of 50$^{''}$ and in the MOS observation of an 
annulus with an inner radius of 30$^{''}$ and and outer radius of 75$^{''}$.
Only events with PATTERN
$\leq$4 for the PN and $\leq$12 for the MOS were selected for
the spectral analysis
with quality parameter FLAG=0. The count rates in all three EPIC detectors were
low enough
(see \S\,\ref{variability}) that the observations were not affected by
pile-ups. Because it was a serendipitous observation, the source was not
observed on CCD \#1 in the MOS cameras, but instead on CCD \#4 and \#6 on MOS-1
and MOS-2, respectively. The position on the EPIC PN was on CCD \#4 close to the
CAMEX.

The XMM data were reduced by using the XMM-Newton Science Analysis Software
(XMMSAS) version 5.4.1 and the X-ray spectra were analyzed by
XSPEC 11.2.0. The 
spectra were grouped by GRPPHA 3.0.0 in bins of at least 15
counts per bin. The response matrices and axillary response files were created
for the XMM
observations by the XMMSAS tasks {\it rmfgen} and {\it arfgen}.

For the sake of completeness and comparison with the XMM data, we retrieved and
reanalyzed the ROSAT PSPC data as well. Details of the ROSAT observations are
given below.

\subsection{ROSAT PSPC observations}

Mkn 1239 was observed twice by ROSAT, first during the RASS \citep{rush96a}
and second in a
pointed PSPC observation \citep{bol96,gru98b, pfe01}.
During the RASS coverage on 1990-11-13 to 1990-11-14
the source was observed for a total of 418s and
the pointed
ROSAT PSPC observation was performed on 1992-11-08 between UT 05:53 - 15:58
for a total of 9043 s (ROR 700908p, \citet{pfe01}). The source was on-axis. 
Source photons
were collected in a circle with a radius of 100$^{''}$ and the background
photons in a circular region close by with a radius of 200$^{''}$.

The ROSAT data were analyzed by both, the
Extended X-ray Scientific Analysis System (EXSAS, \citet{zim98})
version 01APR and XSPEC 11.2.0..
For the count-rate conversions between different X-ray missions,
PIMMS 3.2 was used.

\section{\label{results} Results}

\subsection{\label{variability} X-ray variability}

\begin{figure}[h]
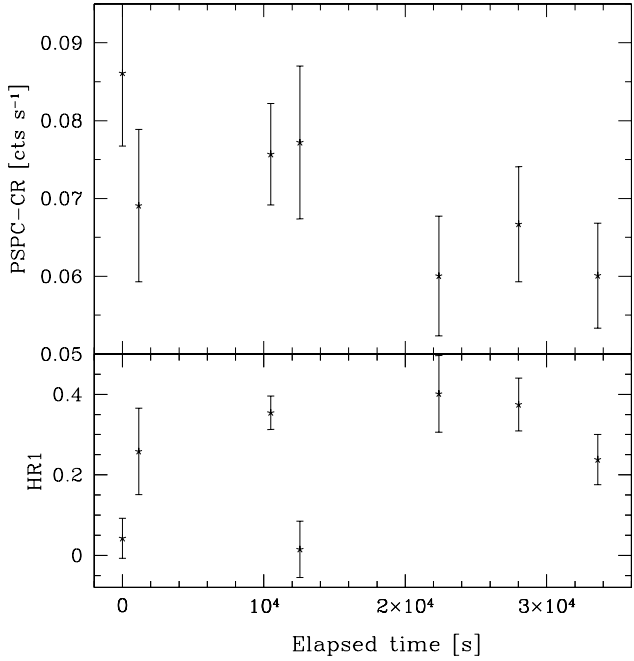

\clipfig{DGrupe.fig1}{84}{13}{60}{180}{240}
\caption[ ]{\label{mkn1239_obi_plot} Light curve of the pointed ROSAT PSPC
observation of Mkn 1239. The upper panel shows the count rate vs. elapsed time
and the lower panel displays the variability of the hardness ratio.
}
\end{figure}

During its RASS and its pointed ROSAT PSPC observations Mkn 1239 had  mean
count rates of 0.054\pl0.014\cts~ and 0.069\pl0.003\cts,
respectively (\citet{pfe01}), suggesting no significant variability in the two
years between the two ROSAT observations. On the other hand, the
light curve of the pointed PSPC observation (Figure\,\ref{mkn1239_obi_plot})
suggests some variability by $\approx$ 25\% during the 10h coverage. 
The PSPC hardness ratios\footnote{HR=(H-S)/(H+S) 
with S counts in the 0.1-0.4 keV range and H=0.5-2.0 keV} were 0.81\pl0.18 and
0.29\pl0.10 during the RASS and pointed PSPC observations suggesting a change in
the X-ray spectrum.
The lower
panel of Figure\,\ref{mkn1239_obi_plot} displays the hardness ratio lightcurve
of the pointed PSPC observation.
It suggests some spectral variability, with the spectrum becoming softer with
increasing count rate.

The mean count rates measured from the EPIC PN and MOS-1 and MOS-2
observations were 0.21\pl0.01\cts, 0.053\pl0.003\cts, and 
0.057\pl0.003\cts, respectively.
No significant variability was detected
during the 5ks and 9.4 ks PN and MOS coverages. 
Using PIMMS and the best-fit single powerlaw in the 0.2-2.0 keV range
with Galactic absorption as given in
Table\,\ref{mkn1239_spec_xmm_rosat} (\S\,\ref{xspectra})
the PN and MOS count rates convert into
0.016 PSPC \cts and 0.019 PSPC \cts~ suggesting a long-term variability by factors
of 3-4 in the 9 and 11 years between the XMM observation and the pointed ROSAT and
RASS observations. 

Converting the single powerlaw spectrum with Galactic
absorption to the PN data as
given in Table\,\ref{mkn1239_spec_xmm_rosat} into a PSPC hardness ratio results in
HR=0.58, suggesting that the source has become harder compared to the pointed
ROSAT PSPC observation. It agrees with the findings for the pointed PSPC
lightcurve (Figure\,\ref{mkn1239_obi_plot}) that the hardness ration increases 
 with decreasing countrate.

\subsection{\label{xspectra} Spectral analysis}

\begin{figure}[t]
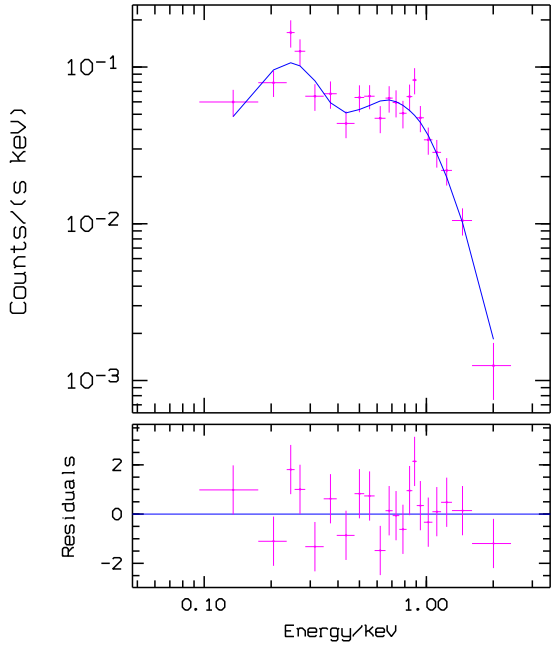

\clipfig{DGrupe.fig2}{77}{15}{90}{142}{182}
\clipfig{DGrupe.fig2}{77}{15}{15}{142}{65}
\caption[ ]{\label{pspc_spec}
Single Power law fit to the pointed PSPC spectrum of Mkn 1239 with absorption at
z=0.0. 
}
\end{figure}

The number of photons collected during the RASS coverage was only 22,
not sufficient enough to perform spectral fits to the data. However, the 9 ks 
pointed PSPC observation was long enough to collect 625 source photons that
allow a spectral analysis. Figure\,\ref{pspc_spec} displays a single power law fit
with neutral absorption at z=0 to the PSPC data of Mkn 1239.
Table\,\ref{mkn1239_spec_xmm_rosat} lists the
results of spectral fits to the ROSAT PSPC. The data are well-fitted by a single
powerlaw with Galactic and intrinsic absorption.
The X-ray spectral index is
\ax=2.74\pl0.27. 
This slope agrees in the 0.1-1.8 keV energy range with the results by using
EXSAS which give \ax=2.79\pl0.30 with an $N_{\rm H}=7.76\times 10^{20}$\cm.
EXSAS has been used for the analysis of the soft X-ray selected AGN sample of
\citet{gru01} and \citet{gru03a}.
The fit to the PSPC data also shows that there is excess absorption above the
Galactic value (\nh=4.03$\times~10^{20}$\cm; \citet{dic90}).
For consistency reasons (see discussion \S\,\ref{dis_slope}) we also fitted an
absorbed powerlaw to the ROSAT PSPC data in the 0.1-2.4 keV range, which results
in an X-ray index \ax=2.97\pl0.30.

\begin{figure}[h]
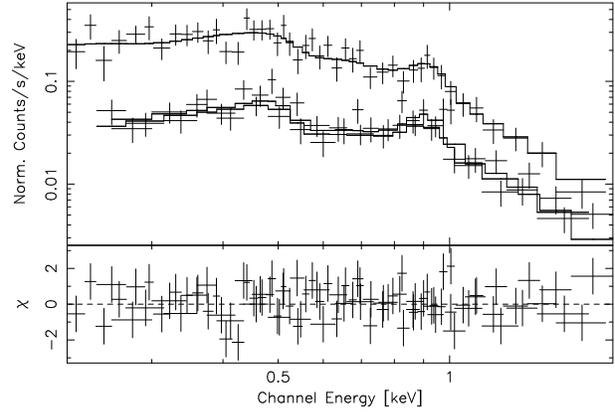

\clipfig{DGrupe.fig3}{84}{03}{06}{250}{170}
\caption[ ]{\label{mkn1239_xmm_02_20} Powerlaw fit with Galactic absorption and
Gaussian line at 0.9 keV to the XMM EPIC PN and MOS spectra in the 0.2-2.0 keV
energy range (Table\,\ref{mkn1239_spec_xmm_rosat}).
}
\end{figure}

A single powerlaw fit to the XMM EPIC PN and MOS data in the 0.2-2.0 keV ROSAT
PSPC energy range confirms the steep X-ray spectrum with \ax=2.95\pl0.24
(Table\,\ref{mkn1239_spec_xmm_rosat}). The fit shows strong residuals around 0.9
keV. This spectral feature can be fitted by a single Gaussian emission line
with an equivalent width of EW=120 eV 
which significantly improves the fit. 
Fig.\,\ref{mkn1239_xmm_02_20} displays this fit to the XMM PN and MOS data.
Even though the PN has the best energy resolution close to the 
CAMEX at 0.9 keV (c.f. \citet{ehle03})
the line is not resolved. The fit to the MOS data results in equivalent widths of
150 eV and 110 eV for the MOS1 and MOS2, respectively. Theoretically, the energy
resolution of the MOS cameras is better than the PN. However, because the source
is rather faint, the MOS data had to be rebinned which reduces the resolution
and the line is not resolved either. 
The line seems to be real and not an effect of the detectors or a result of
mis-calibration, because i) no features in the EPIC detectors around this energy
are known, ii)
it can
be seen in all three XMM detectors, iii) it can
also be seen in the residuals of the ROSAT PSPC spectrum 
(Fig.\,\ref{pspc_spec}), and iv) there are emission lines around this
energy that have been found in the X-ray spectral of several AGN (see discussion
\S\,\ref{dis_eline}).

\begin{figure}[h]
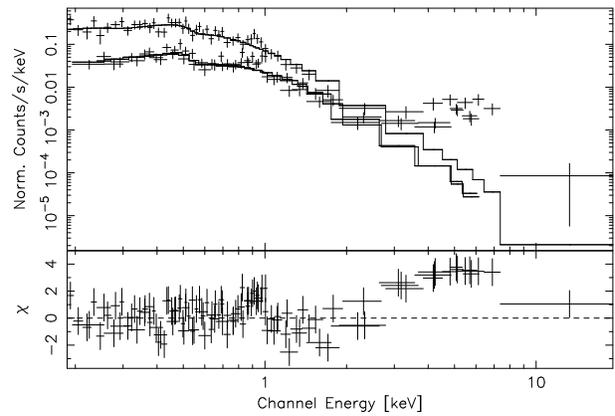

%\clipfig{/home/halley/dgrupe/ps/mkn1239_zwapo_xmm_fit}{84}{03}{06}{250}{170}
\clipfig{DGrupe.fig4}{84}{03}{06}{250}{170}
\caption[ ]{\label{mkn1239_xmm_spec} Power law fit with Galactic and intrinsic 
absorption  fit to the
XMM EPIC PN and MOS data of Mkn 1239. The plot shows that the simple powerlaw
that fits the soft energy range, does not fit the entire spectrum.
}
\end{figure}

\begin{figure}[h]
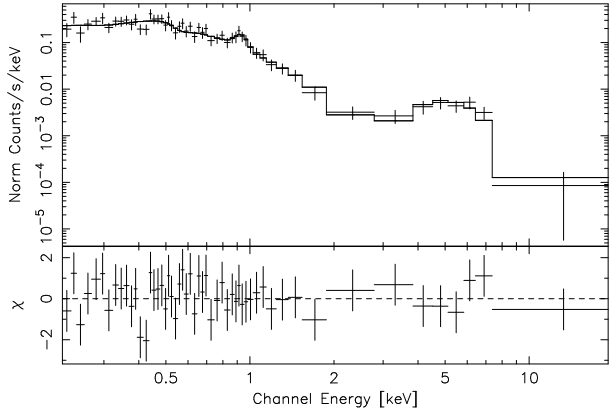

\clipfig{DGrupe.fig5}{84}{03}{06}{250}{170}
\caption[ ]{\label{mkn1239_pn_spec_all} Spectral fit with a powerlaw model
and two intrinsics absorbers and a Gaussian line to the PN data of Mkn 1239
(see Table\,\ref{mkn1239_spec_xmm}).
}
\end{figure}

\begin{figure}[h]
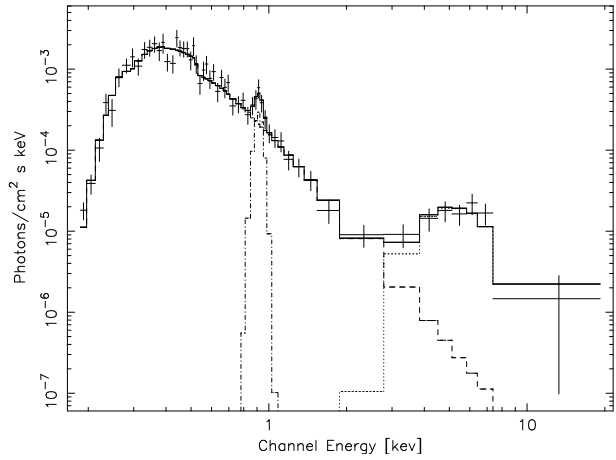

\clipfig{DGrupe.fig6}{84}{03}{03}{250}{190}
\caption[ ]{\label{mkn1239_pn_spec_uf} Unfolded spectrum
 with a powerlaw model
and two intrinsics absorbers and a Gaussian line to the PN data of Mkn 1239
(see Figure\,\ref{mkn1239_pn_spec_all}; Table\,\ref{mkn1239_spec_xmm}).
}
\end{figure}

Even though the 0.2-2.0 keV range seems to be well-fitted by an absorbed
powerlaw model with a Gaussian emission line, the spectrum becomes more
complicated when seen in the 0.2-12 keV range as displayed in
Figure\,\ref{mkn1239_xmm_spec}. A simple absorbed powerlaw model fails to fit
the data above  2 keV. The spectral fit still results in 
a steep X-ray spectral index
\ax$\approx$2.7, but the fit is not acceptable
(Table\,\ref{mkn1239_spec_xmm}). Even though it results in an acceptable fit,
an approach with a broken powerlaw model fails because it needs an
un-physically flat hard X-ray component with \ax=--2.1.
The high degree of optical polarization in Mkn
1239 \citep{good89, brindle90} suggests that part of the observed emission is
scattered while another part of the direct light path is highly absorbed.
Having this in mind, we fitted a powerlaw model with two absorbers to the data.
The result of this fit is shown 
in Figure\,\ref{mkn1239_pn_spec_all}.
This
improves the fit to $\chi^2/\nu$=38.3/50. Figure\,\ref{mkn1239_pn_spec_uf}
displays the unfolded spectrum of this fit showing the soft and hard X-ray
continuum components and the Gaussian line at 0.91 keV.
This result is somewhat similar what
has been seen in the Seyfert 1.5 galaxy Mkn 6 \citep{feldmeier99, immler03}. 
The two absorber model is identical to a partial
covering model  in which the partially covered
continuum component passes a second absorber.
It can be interpreted either as a leaky absorber or an additional line of sight
thought a scattering medium (see \S\,\ref{dis_pcf}).
Partial covering models have been successfully fitted to the XMM data of NLS1s,
such as 1H0707--495 \citep{bol02} and IRAS 13224--3809 \citep{bol03}.

As shown in Figure\,\ref{mkn1239_pn_spec_all} there are
still some residuals around 6.5 keV suggesting the presence of a Fe K$\alpha$
line complex. Adding an additional Gaussian to the data results in a line with
an equivalent width EW=1.3keV, but the quality of the data does not allow to
constrain any of the line parameters.

\subsection{\label{sed} Spectral Energy Distribution}

\begin{figure}[h]
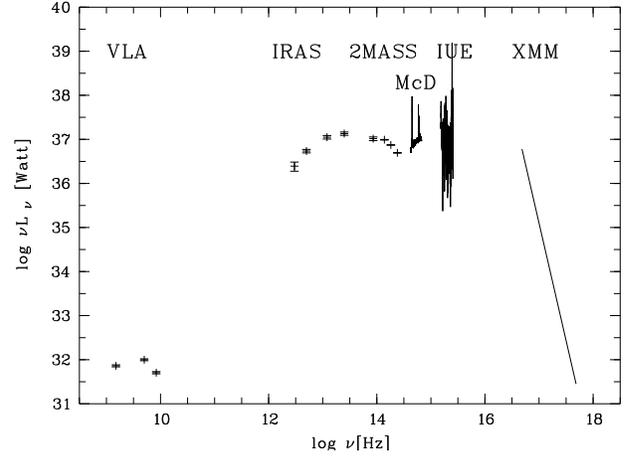

\clipfig{DGrupe.fig7}{84}{18}{08}{275}{195}
\caption[ ]{\label{mkn1239_sed} Spectral Energy Distribution of Mkn 1239. The
NIR, optical and UV data are corrected for reddening. The XMM data are
represented by an unabsorbed powerlaw of the EPIC PN observation in the 0.2-2.0
keV range as given in Tab. \ref{mkn1239_spec_xmm_rosat}.
}
\end{figure}

Fig. \ref{mkn1239_sed} displays the Spectral Energy Distribution (SED) of Mkn
1239. Radio observations of Mkn 1239 have been published by 
\citet{ulv95,rush96b}, and \citet{thean00} showing a compact radio source.
In the plot we used the values of \citet{ulv95}. The far
infrared luminosities were derived from the IRAS point source catalogue and the
near infrared data were taken from the 2 Micron All-Sky Survey (2MASS). The
optical spectrum was observed in 1997 at McDonald Observatory \citep{gru98b}
and the UV spectrum was from an IUE observation of Mkn 1239 (SWP 33659) derived
from the IUE archive at VILSPA.
The X-ray
data are represented by the un-absorbed  powerlaw model that was fitted to
the EPIC PN data in the 0.2-2.0 keV energy range
(Table\,\ref{mkn1239_spec_xmm_rosat}).

One of the crucial points in determining
the SED of this highly reddened AGN is de-reddening.
For the optical and UV data we used the values given in Table 1 in
\citet{gas03}. For the UV we used a fixed ratio $A_{\lambda}/A_V$=1.44 and for
the optical wavelength range we determined $A_{\lambda}/A_V~=~1.9762 - 1.7724
\times 10^{-4}~\times~\lambda$ with $\lambda$ in units of \AA. For the 2MASS
data we used $A_{\lambda}/A_V~=~0.68\times(1/\lambda-0.35)$ with $\lambda$ in
units of $\mu$m (\citet{ward80}).

The 6cm radio flux density which is used for the definition of
the radio loudness \citep{kel89} is 19.49\pl1.05 mJy and 25.9\pl1.3 mJy
(\citet{ulv95} and \citet{rush96b}, respectively). Using the definition of
\citet{kel89} with 
R\footnote{Radio loudness R=$F_{\rm 5~GHz}/F_{\rm 4400\AA}>$10} 
for radio-loud sources and the 
de-reddened flux density at 4400\AA~ the radio loudness becomes R=5.3 and 7.0,
respectively. This would make Mkn 1239 a radio-quiet source. Nevertheless it is
a borderline object between radio-loud and radio-quiet.

We determined the X-ray loudness \aox\footnote{\aox~
is the slope of a hypothetical power-law from 2500 \AA~ to 2 keV;
\aox=0.384$\log~(L_{2500}/L_{2keV})$}  using
the unabsorbed monochromatic luminosities $L_{\rm 2500\AA}$ and $L_{\rm
2keV}$ which were measured directly from the SED given in Fig.
\ref{mkn1239_sed}. This results in \aox=1.50. This agrees almost perfectly with
the value of \aox~given by \citet{yua98} for a source of a redshift z$<$0.2 and
and optical luminosity density log $l_{\rm opt}$=23 [W Hz$^{-1}$] (30 [ergs
s$^{-1}$ Hz$^{-1}$]).

From the best-fit two absorber powerlaw model given in 
table\,\ref{mkn1239_spec_xmm} we estimated the unabsorbed 
0.2-2.0 rest-frame X-ray luminosity
log $L_{\rm 0.2-2.0 keV}$ = 37.27 [W] (44.27 [ergs s$^{-1}$])
and the 0.2-12 keV rest-frame luminosity 
log $L_{\rm 0.2-12.0 keV}$ = 37.29 [W] (44.29 [ergs s$^{-1}$])
which makes Mkn 1239 one of the
high-luminous NLS1s compared with the sample of \citet{gru03a}. The contribution
of the soft, scattered component is log $L_{\rm 0.2-12 keV}$(soft) = 35.00 [W]
(42.00 [ergs s$^{-1}$]).
This is about 1/200th of the unabsorbed total X-ray luminosity in the 0.2-12 keV
band and agrees well with the covering fraction of 0.995\% found when a partial
covering absorption model was used to fit the data.
The monochromatic luminosity at 5100\AA~ is 
log $\lambda L_{5100}$=36.90 [W] (43.90 [ergs s$^{-1}$]). With
a FWHM(H$\beta$)=1050\kms~and the relations given in \citet{kas00} we estimated
the mass of the central black hole to $M_{\rm BH}~=~5\times 10^6~M_{\odot}$.
This converts to an Eddington luminosity of 
log $L_{\rm Edd}$ = 37.7 [W] (44.7 [ergs s$^{-1}$]). From
the SED the bolometric luminosity was estimated to log $L_{\rm bol}~\approx$
38.0 [W] by using a powerlaw with exponential cutoff plus 
soft X-ray powerlaw with neutral absorption to
model the Big Blue Bump emission  as described by \citet{gru03a}. 
This suggests that Mkn 1239
has an Eddington ratio $L/L_{\rm Edd}$ of about 2. 

\section{\label{discuss} Discussion}

\subsection{\label{dis_slope} The steep X-ray slope \ax}

\begin{figure}[h]
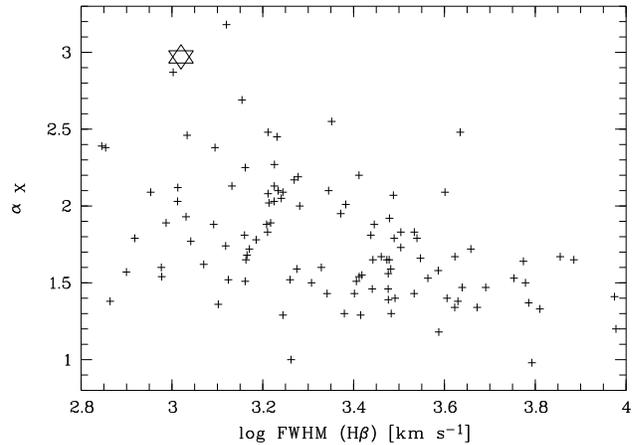

\clipfig{DGrupe.fig8}{84}{20}{12}{275}{192}
\caption[ ]{\label{mkn1239_fwhb_ax} FWHM(H$\beta$) - \ax~ diagram of the soft
X-ray selected AGN sample of \citet{gru03a} and \citet{gru03b}. The position of
Mkn 1239 is marked by the star.
}
\end{figure}

The soft X-ray spectrum with \ax$\approx$3.0
is unusually steep even for a NLS1, but not uncommon (e.g. 
\citet{bol96} and \citet{gru01}). Its black hole mass of
5$\times 10^6~M_{\odot}$
and Eddington accretion ratio $L/L_{\rm Edd}$ of about 2-3
are similar to what has been found for other NLS1s (e.g. \citet{gru03a} and
\citet{gru03b}). With an \ax=2.97 and 
a FWHM(H$\beta$)=1050\pl150 \kms~\citep{gru98b}, Mkn 1239 is one of the 
extreme sources in a FWHM(H$\beta$)-\ax-diagram
(c.f. e.g. \citet{bol96, gru99, gru03b, wil02}). 
Figure\,\ref{mkn1239_fwhb_ax} displays the position of Mkn 1239
(marked as the large  star)
in the FWHM(H$\beta$) -
\ax~diagram of the complete soft X-ray selected sample of \citet{gru03a} and
\citet{gru03b}. The value taken for this plot was \ax=2.97 to be consistent with
the other objects for which EXSAS had been used for the ROSAT data analysis.
Please note, that Mkn 1239 is the only NLS1 in this plot that
was not selected by X-rays.

\subsection{\label{dis_pcf} Partial covering and its relation to optical polarization}

The degree of optical polarization  is wavelength dependent
and increases with decreasing wavelength \citep{good89, brindle90} 
suggesting two main
light paths between the continuum source and the observer, one direct, highly
reddened absorbed light path
and one indirect scattered and less-absorbed light path. The XMM data seem
to confirm this model. The best-fit model to the PN data contains a powerlaw
with two absorbers (Table\,\ref{mkn1239_spec_xmm}). 
We interpret the soft component not as direct light from a leaky absorber
situation, but rather as a scattered component. This interpretation is motivated
by the high degree of optical poarization.
This picture is somewhat similar to the model proposed by
\citet{feldmeier99} based on ASCA observations of Mkn 6 which was confirmed by
the XMM observation \citep{immler03}. 

\citet{smith04} have noticed that the polarization in
the H$\alpha$ line shows a minimum of polarization in the blue wing, while the
polarization in the red wing shows a peak. Therefore, \citet{smith04}
concluded that the
scattering medium is part of the nuclear outflow. A strong outflow is expected
in a sources with high $L/L_{\rm Edd}$ (e.g. \citet{king03}) like Mkn 1239.

\subsection{\label{dis_eline} The strong NeIX emission line}

One of our main findings in the X-ray spectrum of Mkn 1239 is a strong emission
feature around 0.9 keV.
Given the energy of the feature, the most
obvious identifications would be the Ne IX triplet.
However, the strength of the line, and the lack of any other detectable
emission features, is surprising. 
In the first case, the absence of the oxygen triplet
is unexpected.  For solar metal abundances that lines are expected
to be stronger than the Neon features. Taking the line parameters of 
the 0.91 keV feature, we found
an upper limit for the 22\AA~ OVII triplet of an equivalent widths EW=34\AA.
 An identification of the line with Neon would thus
lead us back to speculations about a neon/oxygen overabundance first
discussed in the context of PG1404+226 \citep{kom98, ulrich99}.
In the latter case, features were seen in absorption, while we detect an
emission-line in Mkn  1239. Note that Ne and/or Fe-L lines are 
relatively stronger to oxygen in AGN with
steep X-ray spectra (c.f. \citet{nic99}).
A possible origin of that line then is the ionized
medium not located along the line-of-sight, thus seen
in emission. 

The presence of He-like ions like NeIX in emission suggests a hot
plasma with temperature of several million K
(e.g. \citet{anil82, porquet00, porquet01} and \citet{ness03}
 and references
therein). These type of emission lines have been found in the X-ray spectra of
hot stars (e.g. \citet{audard01, stelzer02} and \citet{ness03}), and have also 
been reported for a few AGN,
e.g. NGC 3783 \citep{kas02, kron03}, 
NGC 1068 \citep{kin02}, NGC 5548 \citep{kaastra00}
and NGC 4151 \citep{ogle00}.
The 0.91 keV line
found in Mkn 1239 from the XMM data show an EW$\approx$120 eV.
\citet{comastri98} reported of a strong
NeIX line in the ASCA spectrum of the Seyfert 2
galaxy NGC 4507 with a similar equivalent width as found in Mkn 1239. The
continuum spectrum of NGC 4507 is somewhat similar to Mkn 1239 in showing a
highly absorbed hard X-ray component and a less absorbed soft component. 
In both cases, emission lines are strongly detected because the direct light
 is supressed.
Theoretically the presence of an He-like ion triplet provides a powerful tool to
determine parameters of the plasma such as electron temperature and density
(e.g. \citet{netzer96})
Unfortunately in the case of Mkn 1239 the source is too faint to be explored by
XMM's RGS. Much more powerful observatories such as XEUS and CON-X have to be
used to observe this source with high resolution X-ray spectrometers.

%Instead of strong signs of a warm absorber we found a strong emission line at
%0.92 in the observed frame, most likely the NeIX triplet.

\subsection{X-ray variability}

The variability of a factor of about 4 between the pointed ROSAT PSPC and the
XMM observation about 9 years later cannot be explained only by a change in the
intrinsic absorption column. Using the  changes in the absorption columns of the
ROSAT PSPC and XMM observations as given in Table\ref{mkn1239_spec_xmm_rosat}
would only cause a variability by a factor of about 2. 
This means that the source has to be variable
intrinsically. Changes in the observed X-ray flux by factors of 3-4 
are not uncommon among AGN (e.g. \citet{lei99, gru01} and references therein),
especially NLS1s (e.g. \citet{bol96}).
The fits to the PSPC and XMM data in the 0.2-2.0 keV energy
range do not suggest any significant change in the soft X-ray slope
(Table\,\ref{mkn1239_spec_xmm_rosat}). 
From the
current data sets, the fits are consistent with a spectrum that goes just up and
down without any significant changes of its spectral slope \ax.

\section{Conclusions}
We have studied the X-ray spectra of Mkn 1239 in a serendipitous observation by
XMM and our three main results are:  
\begin{enumerate}
\item The 0.2-2.0 keV
soft X-ray spectrum of the source is very steep with \ax$\approx$3.0 confirming
previous results from ROSAT. 
\item The 0.2-12 keV EPIC PN
X-ray continuum spectrum is best-fitted by a powerlaw with \ax=2.92 and
 two intrinsic absorbers with $N_{\rm H}=6\times10^{20}$\cm and
 3$\times10^{23}$\cm, supporting the optical spectropolarimetry results by
 \citet{good89} of having a direct highly absorbed light path and an indirect
 scattered light path which is less absorbed.
\item The X-ray spectrum shows an emission line feature
around 0.91 keV (observed frame) which is most likely the NeIX triplet. 
\item The non-detection of the OVII triplet suggests a super-solar Ne/O ratio.
\item To finally resolve the 0.91 keV feature, longer
observations with XMM or even more powerful future X-ray missions are needed.
\end{enumerate}

\acknowledgments

We would like to thank Michael Freyberg and Frank Haberl for intensive
discussions about the EPIC PN and MOS calibration. Many thanks also to Anil
Pradhan for discussion on Helium-like ions and Joe Shields for helpful comments
and discussions. 
This research 
has made use of the NASA/IPAC Extra-galactic
Database (NED) which is operated by the Jet Propulsion Laboratory, Caltech,
under contract with the National Aeronautics and Space Administration. 
The ROSAT project is supported by the Bundesministerium f\"ur Bildung
und  Forschung (BMBF/DLR) and the Max-Planck-Society. This work was supported in
part by NASA grant NAG5-9937.


\begin{deluxetable}{llccccr}
\tabletypesize{\scriptsize}
\tablecaption{Spectral Fit parameters to the EPIC PN and MOS and ROSAT PSPC
data of Mkn 1239 in the ROSAT PSPC 0.1-2.0 keV energy range.
 \label{mkn1239_spec_xmm_rosat}}
\tablewidth{0pt}
\tablehead{
& &  \colhead{$N_{\rm H, gal}$} & \colhead{$N_{\rm H, intr}$}  & 
&  \colhead{EW(NeIX)} \\
\colhead{Detector} &
\colhead{\rb{XSPEC Model}} & \colhead{10$^{20}$\cm}  & \colhead{10$^{20}$\cm}   &
\colhead{\rb{$\alpha_{\rm X}$}} & \colhead{\AA} &
\colhead{\rb{$\chi^2$ (DOF)}} 
} 
\startdata
PSPC\tablenotemark{1} & wa po\tablenotemark{3} & 7.77\pl0.95 & --- & 
2.73\pl0.27 & --- & 23.0 (28) \\
& wa zwa po\tablenotemark{4} & 4.03 (fix) & 3.91\pl1.00 & 2.74\pl0.29 & 
--- & 23.0 (28) \\ \\
PN\tablenotemark{2} & wa po\tablenotemark{3} &  10.94\pl2.07 & --- & 
2.95\pl0.24 & --- & 47.7 (45) \\
& wa zwa po\tablenotemark{4} &  4.03 (fix) & 7.47\pl2.21 & 2.97\pl0.25 & 
--- & 47.5 (45) \\
& wa po gaus\tablenotemark{5} & 9.75\pl1.95 & --- & 2.94\pl0.24 & 120 & 34.7
(42) \\ \\
MOS-1+MOS-2\tablenotemark{2} & wa po\tablenotemark{3} & 16.50\pl3.99 & --- &
3.34\pl0.36 & --- & 53.0 (42)  \\
& wa zwa po\tablenotemark{4} & 4.03 (fix) & 14.15\pl4.45 & 3.44\pl0.38 
& --- & 52.6 (42) \\
& wa po gaus\tablenotemark{5} & 12.02\pl3.87 & --- & 3.04\pl0.35 & 153, 104 &
37.4 (37) \\ \\
PN+MOS-1+MOS-2\tablenotemark{2} & wa po\tablenotemark{3} & 13.60\pl2.15 & --- &
3.14\pl0.21 & --- & 98.2 (86) \\
& wa zwa po\tablenotemark{4} & 4.03 (fix) & 10.07\pl2.41 & 3.21\pl0.23 & 
--- & 97.5 (86) \\
& wa po gaus\tablenotemark{5} & 10.41\pl1.59 & --- & 2.94\pl0.18 & 120, 150, 110
& 74.4 (84) 
\enddata

\tablenotetext{1}{ROSAT PSPC, observed energy range 0.1-1.8 keV}
\tablenotetext{2}{EPIC PN, MOS-1 and MOS-2observed energy range 0.2-2.0 keV}
\tablenotetext{3}{galactic absorption and powerlaw model}
\tablenotetext{4}{galactic absorption, redshifted neutral absorption at z=0.02
and powerlaw}
\tablenotetext{5}{powerlaw with Galactic absorption and Gaussian line at
$\approx$0.92 keV}
\end{deluxetable}

\begin{deluxetable}{lcccc}
\tabletypesize{\scriptsize}
\tablecaption{Spectral Fit parameters to the EPIC PN 0.2-12 keV
data of Mkn 1239. The Galactic \nh~is fixed for all models to 4.03 $10^{20}$\cm.
\label{mkn1239_spec_xmm}}
\tablewidth{0pt}
\tablehead{
& \multicolumn{3}{c}{Models} \\ 
\colhead{Parameter} & \colhead{ZWA PO\tablenotemark{1}} & 
\colhead{ZWA(PO+GA)\tablenotemark{2}} &
\colhead{ZWA(PO+GA)+ZWA(PO)\tablenotemark{3}}   
} 
\startdata
Absorber 1, \nh~[$10^{20}$\cm] & 5.76\pl1.82 & 5.06\pl1.81 &  
6.21\pl2.01 \\
Absorber 2, \nh~[$10^{22}$\cm] & --- & ---  & 33.24\pl13.46 \\
Spectral index \ax & 2.75\pl0.20 & 2.75\pl0.20 &  
2.92\pl0.23 \\
Gauss line, E [keV] & --- & 0.913\pl0.022 & 
0.913 (fix) \\
Gauss line, $\sigma$ [eV] & --- & 32.4$^{+61.0}_{-32.4}$ & 32.4 (fix)  \\
Gauss line, EW [eV] & --- & 110 &  110 (fix) \\
$\chi^2/\nu$ & 111.9 (53) & 99.6 (50) &  38.3 (50)
\enddata

\tablenotetext{1}{XSPEC model ZWA PO, intrinsic absorption by neutral elements
and powerlaw}
\tablenotetext{2}{XSPEC model ZWA(PO+GAUS), intrinsic absorption with powerlaw
and Gaussian emission line} 
\tablenotetext{3}{XSPEC model ZWA (PO+GAUS) + ZWA (PO), Two-component intrinsic
absorption with powerlaw and Gaussian line}
\end{deluxetable}


\begin{thebibliography}{}
\bibitem[Audard et al. (2001)]{audard01} Audard, M., Behar, E., G\"udel, M.,
Raassen, Porquet, D., Mewe, R., Foley, C.R., \& Bronage, G.E., 2001, \aap, L329
\bibitem[Beuermann et al. (1999)]{beu99} Beuermann, K., Thomas, H.-C., Reinsch,
K., et al., 1999, \aap, 347, 47
\bibitem[Boller et al. (1996)]{bol96} Boller, T., Brandt, W.N., \& Fink, H.H.,
1996, \aap, 305, 53
\bibitem[Boller et al. (2002)]{bol02} Boller, T., Fabian, A.C., Sunyaev, R., et
al., 2002, \mnras, 329, 1
\bibitem[Boller et al. (2003)]{bol03} Boller, T., Tanaka, Y., Fabian, A., et
al., 2003, \mnras, 343, 89
\bibitem[Boroson \& Green (1992)]{bor92} Boroson, T.A., \& Green, R.F., 1992,
\apjs, 80, 109
\bibitem[Boroson (2002)]{bor02} Boroson, T.A., 2002, \apj, 565, 78
%\bibitem[Brandt et al (1994)]{bra94} Brandt W.N., Fabian A.C., Nandra K.,
% Reynolds C.S., Brinkmann W., 1994, \mnras, 271, 958
\bibitem[Brindle et al. (1990)]{brindle90} Brindle, C., Hough, J.H., Bailey,
J.A., Axon, D.J., Ward, M.J., Sparks, W.B., \& McLean, I.S., 1990, \mnras, 244,
577
\bibitem[Comastri, et al. (1998)]{comastri98} Comastri, A., Vignali, C., Cappi,
M., Matt, G., Audano, R., Awaki, H., \& Ueno, S., 1998, \mnras, 295, 443
\bibitem[Dickey \& Lockman (1990)]{dic90} Dickey, J.M., \& Lockman, F.J., 1990,
\araa, 28, 215
\bibitem[Ehle et al. (2003)]{ehle03} Ehle, M., Breitfellner, M., Gonzales
Riestra, R., et al., 2003, XMM-Newton Users' Handbook, Issue 2.1
\bibitem[Feldmeier et al (1999)]{feldmeier99} Feldmeier, J.J., Brandt, W.N.,
Elvis, M., Fabian, A.C., Iwasawa, K., \& Mathur, S., 1999, \apj, 510, 167
\bibitem[Gaskell et al. (2003)]{gas03} Gaskell, C.M., Goosmann, R.W., Antonucci,
R.R.J., \& Whysong, D., 2003, ApJ submitted, astro-ph/0309595
\bibitem[Goodrich (1989)]{good89} Goodrich, R.W., 1989, \apj, 342, 224
\bibitem[Grupe(1996)]{gru96} Grupe, D., 1996, PhD Thesis, Universit\"at
G\"ottingen
\bibitem[Grupe (2004)]{gru03b} Grupe, D., 2004, \aj, accepted, astro-ph/0401167
\bibitem[Grupe et al. (1998a)]{gru98a} Grupe, D., Beuermann, K., Thomas, H.-C.,
Mannheim, K., \& Fink, H.H., 1998, A\&A 330, 25
\bibitem[Grupe et al. (1998b)]{gru98b} Grupe, D., Wills, B.J., Wills, D.,
Beuermann, K., 1998, \aap, 333, 827
\bibitem[Grupe et al.(1999)]{gru99} Grupe, D., Beuermann, K., Mannheim, K.,
\& Thomas, H.-C., 1999, \aap, 350, 805
\bibitem[Grupe et al.(2001)]{gru01} Grupe, D., Thomas, H.-C., \& Beuermann, K.,
2001, \aap, 367, 470 
\bibitem[Grupe et al. (2004a)]{gru03a} Grupe, D., Wills, B.J., Leighly, K.M., \&
Meusinger, H., 2004a, \aj, 127, 156
\bibitem[Immler et al. (2003)]{immler03} Immler, S., Brandt, W.N., Vignali, C.,
Bauer, F.E., Crenshaw, D.M., Feldmeier, J.J., \& Kraemer, S.B., 2003, \aj, 126,
153
\bibitem[Jansen et al. (2001)]{jan01} Jansen, F., Lumb, D., Altieri, B., et al.,
2001, \aap, 365, L1
\bibitem[Kaastra et al. (2000)]{kaastra00} Kaastra, J.S., Mewe, R., Liedahl,
D.A., Komossa, S., \& Brinkmann, A.C., 2000, \aap, 354, L83
\bibitem[Kaspi et al. (2000)]{kas00} Kaspi, S., Smith, P.S., Netzer, H., Moaz,
D., Jannuzi, B.T., \& Giveon, U., 2000, \apj, 533, 631
\bibitem[Kaspi et al. (2002)]{kas02} Kaspi, S., Brandt, W.N., George, I.M., et
al., 2002, \apj, 574, 643
\bibitem[Kellermann et al. (1989)]{kel89} Kellermann, K.I., Sramek, R., Schmidt,
M., Shaffer, D.B., \& Green, R., 1989, \aj, 98, 1195
\bibitem[King \& Pounds (2003)]{king03} King, A.R., \& Pounds, K.A., 2003,
\mnras, 345, 657
\bibitem[Kinkhabwala et al. (2002)]{kin02} Kinkhabwala, A., Sako, M., Behar, E.,
et al., 2002, \apj, 575, 732
\bibitem[Komossa \& Fink (1998)]{kom98} Komossa S., \& Fink H., 
1998, in: Highlights in X-ray astronomy,
   B. Aschenbach \& M.J. Freyberg (eds.), MPE Report 272, 147
%\bibitem[Komossa et al (2001)]{kom01} Komossa S., Burwitz, V., Predehl P., 
%Kaastra J., 2001, in: 
%     The central kpc of starbursts and AGN: the La Palma connection, 
%      J.H. Knapen et al.  (eds), ASP Conf. series 249, 411
\bibitem[Krongold et al. (2003)]{kron03} Krongold, Y., Nicastro, F., Brickhouse,
N.S., Elvis, M., Liedahl, D.A., \& Mathur, S., 2003, \apj, 597, 832
\bibitem[Laor et al. (1994)]{laor94} Laor, A., Fiore, F., Elvis, M., Wilkes,
B.J., \& McDowell, J.C., 1994, \apj, 435, 611
\bibitem[Laor et al. (1997)]{laor97} Laor, A., Fiore, F., Elvis, M., Wilkes,
B.J., \& McDowell, J.C., 1997, \apj, 477, 93
\bibitem[Leighly (1999)]{lei99} Leighly, K.M., 1999, \apjs, 125, 317
\bibitem[Leighly et al. (1997a)]{lei97a} Leighly, K.M., Mushotzky, R.F., Nandra,
K., \& Forster, K., 1997, \apj, 489, L25
\bibitem[Leighly et al. (1997b)]{lei97}  Leighly, K.M., Kay, L.E., Wills, B.J.,
Wills, D., \& Grupe, D., 1997, \apj, 489, L137  
\bibitem[Ness et al. (2003)]{ness03} Ness, J.-U., Brickhouse, N.S., Drake, J.J.,
\& Huenemoerder, D.P., 2003, \apj, 598, 1277
\bibitem[Netzer (1996)]{netzer96} Netzer, H., 1996, \apj, 473, 781
\bibitem[Nicastro et al. (1999)]{nic99} Nicastro F., Fiore F., Matt G., 
1999, \apj, 517, 108
\bibitem[Ogle et al. (2000)]{ogle00} Ogle, P.M., Marshall, H.L., Lee, J.C., \&
Canizares, C.R., 2000, \apj, 545, L81
\bibitem[Osterbrock \& Pogge (1985)]{oster85} Osterbrock, D.E., \& Pogge, R.W.,
1985, \apj, 297, 166
%\bibitem[Otani et al. (1996)]{otani96} Otani C., Kii T., Miya K., 1996,
%in MPE Report 263, H.U. Zimmermann, J. Tr\"umper, H. Yorke (eds), 491
\bibitem[Pfefferkorn et al. (2001)]{pfe01} Pfefferkorn, F., Boller, T., \&
Rafanelli, P., 2001, \aap, 368, 797 
\bibitem[Pfeffermann et al.(1987)]{pfe86} Pfeffermann, E., Briel, U.G.,
Hippmann, H., et al., 1987, SPIE, 733, 519
\bibitem[Porquet \& Dubau (2000)]{porquet00} Porquet, D., \& Dubau, J., 2000,
\aaps, 143, 495
\bibitem[Porquet et al. (2001)]{porquet01} Porquet, D., Mewe, R., Dubau, J.,
Raassen, A.J.J., Kaastra, J.S., 2001, \aap, 376, 1113 
%\bibitem[Pounds et al. (2003)]{pounds03} Pounds, K.A., Reeves, J.N., King, A.R.,
%Page, K.L., O'Brian, P.T., \& Turner, M.J.L., 2003, \mnras, 345, 705
\bibitem[Pradhan (1982)]{anil82} Pradhan, A., 1982, \apj, 263, 477
\bibitem[Rush et al. (1996a)]{rush96a} Rush, B., Malkan, M.A., Fink, H.H., \&
Voges, W, 1996, \apj, 471, 190
\bibitem[Rush et al. (1996b)]{rush96b} Rush, B., Malkan, M.A., \& Edelson, R.A.,
1996, \apj, 473, 130
\bibitem[Schwope et al. (2000)]{schwo00} Schwope, A.D., Hasinger, G., Lehmann,
I., et al., 2000, AN, 321, 1
%\bibitem[Smith et al. (2002)]{smith02} Smith, J.E., Young, S., Robinson, A.,
%Corbett, E.A., Giannuzzo, M.E., Axon, D.J., \& Hough, J.H., 2002, \mnras, 335,%
%773
\bibitem[Smith et al. (2004)]{smith04} Smith, J.E., Robinson, A., Alexander,
D.M., Young, S., Axon, D.J., \& Corbett, E.A., 2004, \mnras, accepted,
astro-ph/0401496
\bibitem[Stelzer et al. (2002)]{stelzer02} Stelzer, B., Burwitz, V., Audard, M.,
et al., (2002), \aap, 392, 585
\bibitem[Str\"uder et al.(2001)]{str01} Str\"uder, L., Briel, U., Dennerl, K.,
et al., 2001, \aap, 365, L18
\bibitem[Sulentic et al. (2000)]{sul00} Sulentic, J.W., Zwitter, T., Marziani,
P., \& Dultzin-Hacyan, D., 2000, \apj, 536, L5
\bibitem[Thean et al. (2000)]{thean00} Thean, A., Pedlar, A., Kukula, M.J.,
Baum, S.A., \& O'Dea, C.P., 2000, \mnras, 314, 573
\bibitem[Thomas et al. (1998)]{tho98} Thomas, H.-C., Beuermann, K., Reinsch, K.,
et al., 1998, \aap, 335, 467
\bibitem[Tr\"umper(1982)]{tru83} Tr\"umper, J., 1982, Adv. Space Res., 4, 241
%\bibitem[Turner et al. (1999)]{turner99} Turner T.J., George I.M., Netzer H., 
%1999, \apj, 526, 52
\bibitem[Turner et al.(2001)]{tur01} Turner, M.J.L., Abbey, A., Arnaud, M., et
al., 2001, \aap, 365, L27
\bibitem[Ulrich et al. (1999)]{ulrich99} Ulrich M.-H., Comastri A., Komossa S., 
Crane P., 1999, \aap, 350, 816
\bibitem[Ulvestad et al. (1995)]{ulv95} Ulvestad, J.S., Antonucci, R.R.J., \&
Goodrich, R.W., 1995, \aj, 109, 81
\bibitem[Vaughan et al. (2001)]{vau01} Vaughan, S., Edelson, R., Warwick, R.S.,
Malkan, M,A., \& Goad, M.R., 2001, \mnras, 327, 673
\bibitem[Voges et al. (1999)]{vog99} Voges, W., Aschenbach, B., Boller, T., et
al., 1999, \aap, 349, 389
\bibitem[Ward et al. (1980)]{ward80} Ward, M., Penston, M.V., Blades, J.C., \&
Turtle, A.J., 1980, \mnras, 193, 563
\bibitem[Williams et al. (2002)]{wil02} Williams, R.J., Pogge, R.W., \& Mathur,
S., 2002, \aj, 124, 3042
\bibitem[Wills et al. (1992)]{wills92} Wills, B.J., Wills, D., Evans, N.J.,
Natta, A., Thompson, K.L., Breger, M., \& Sitko, M.L., 1992, \apj, 400, 96
\bibitem[York et al. (2000)]{york00} York et al., 2000, \aj, 120, 1579
\bibitem[Yuan et al. (1998)]{yua98} Yuan, W., Brinkmann, W., Siebert, J., Voges,
W., 1998, \aap, 330, 108
\bibitem[Yuan \& Wills (2003)]{yuan03} Yuan, M.J., \& Wills, B.J., 2003, \apj,
593, L11
\bibitem[Zimmermann et al.(1998)]{zim98} Zimmermann, U., Boese, G., Becker, W.,
et al., 1998, 'EXSAS User's Guide', MPE report 
(http://wave.xray.mpe.mpg.de/exsas/users-guide)
\end{thebibliography}
\end{document}